\documentclass[aps,floatfix,showpacs,twocolumn, preprintnumbers,numbersect]{revtex4-1}
\usepackage[utf8]{inputenc}

\usepackage{bm}
\usepackage{graphicx,epsfig}
\usepackage{amsfonts,amssymb,amsmath}
\usepackage{braket}
\usepackage{slashed}

\usepackage{float}
\usepackage{multirow}
\usepackage{booktabs}
\usepackage{dcolumn}     

\usepackage{caption}
\usepackage{subcaption}

\usepackage{xcolor}
\usepackage{soul}

\usepackage{hyperref}
\hypersetup{colorlinks,urlcolor=blue,allcolors=blue}

\begin{document}
\title{Bayesian analysis of proton-proton fusion in chiral effective field theory}

\author{V.\ Barlucchi$^{1,2,3}$}
\email{vbarlucc@uni-mainz.de}
\author{A.\ Gnech$^{4,5}$}
\email{agnech@odu.edu}
\author{S.\ Degl'Innocenti$^{2,3}$}
\email{scilla.deglinnocenti@unipi.it}
\author{L.E.\ Marcucci$^{2,3}$}
\email{laura.elisa.marcucci@unipi.it}
\affiliation{
$^{\rm 1}$\mbox{Institut für Kernphysik Mainz, Johannes Gutenberg Universität Mainz, D-55099 Mainz, Germany}\\
$^{\rm 2}$\mbox{Dipartimento di Fisica “E. Fermi”, Universit\`a di Pisa, Pisa I-56127, Italy}
$^3$\mbox{Istituto Nazionale di Fisica Nucleare, Sezione di Pisa, Pisa I-56127, Italy}
$^4$\mbox{Department of Physics, Old Dominion University, Norfolk, VA 23529, USA}\\
$^5$\mbox{Theory Center, Jefferson Lab, Newport News, VA 23610, USA}\\
}
\date{\today}
\begin{abstract}
The astrophysical $S$-factor for the proton-proton fusion is calculated in the low-energy regime for a variety of nuclear interactions and consistent nuclear currents, derived within chiral effective field theory. We estimate, for the first time, the theoretical uncertainty on the $S$-factor due to the truncation of the chiral expansion of the currents using a Bayesian analysis. In order to reach an accuracy at the percent level in the calculation, the electromagnetic potential includes contributions beyond the leading Coulomb interaction, such as two-photon exchange and vacuum polarization. The initial proton-proton state is expanded in partial waves and only the ${}^1S_0$ contribution is included, as it is known that the other partial-waves effects are negligible. The  low-energy constant entering the contact term in the weak axial current operator is calibrated to reproduce the Gamow-Teller matrix element in Tritium $\beta$-decay.  The value $S(0)$ is found to be $S(0)=(4.068 \pm 0.025)\times 10^{-25} \: \text{MeV}\: \text{b}$.
\end{abstract}

\maketitle
\section{Introduction and conclusions}
\label{sec:IntroConcl}
The proton-proton ($pp$) fusion, i.e. the process
\begin{equation}\label{eq:ppreaction}
    p+p\xrightarrow{}d+e^-+\bar\nu_e \, ,
\end{equation}
is the driving reaction of the $pp$ chain, which is the main source of energy production in stars with a  mass of the order of the Sun (see the most recent review of Ref.~\cite{SFIII}). 
The rate of the $pp$ reaction controls the velocity of the full chain and therefore it is crucial to determine with great accuracy and controlled errors its cross section $\sigma(E)$, or, alternatively, the astrophysical $S$-factor $S(E)$, defined as
\begin{equation}{\label{eq:sfactor}}
    S(E)=E\,\sigma(E)\,e^{-2\pi\eta(E)}\ ,
\end{equation}
$E$ being the center-of-mass energy, $\eta(E)=\alpha/v$ the Sommerfeld parameter, with $\alpha\sim 1/137$ the fine-structure constant, $v =\sqrt{E/m}$ the $pp$ relative velocity, with $m$ being the proton mass.  
Unfortunately, the $pp$ cross section in the energy range of interest for the Sun, i.e.\ around the Gamow peak ($E\simeq 6$ keV), is too small to be measured in terrestrial laboratories, and thus must be obtained from theoretical calculations, possibly accompanied by a robust estimate of their uncertainties. 
The subject of this work is a theoretical study of $S(E)$ accompanied by a robustly estimated theoretical error based on the Bayesian analysis of the truncation errors of the theory. As final result, we will provide also the best estimate with error for $S(0)$, its first derivative $S^\prime(0)$ and its second derivative $S^{\prime\prime}(0)$, necessary to perform the Taylor expansion for $S(E)$, i.e. 
\begin{equation}{\label{eq:taylor}}
S(E)=S(0) + S^\prime(0)E 
+\frac{1}{2}S^{\prime\prime}(0)E^2 + 
\cdots \ ,  
\end{equation}
where the dots represent further contributions, usually not needed. The Taylor expansion is typically used in astrophysics, when the full energy dependence is not known.

\subsection{Previous studies and motivations}
\label{subsec:previous}

A historical summary of the theoretical studies of the $pp$ reaction can be found in the series of reviews known as "Solar Fusion" (SF), SFI~\cite{SFI}, SFII~\cite{SFII} and most recently SFIII~\cite{SFIII}. These reviews also provided the best estimates at that time for $S(0)$, and, starting from SFII, also for its first and second derivatives at zero energy, $S^\prime(0)$ and $S^{\prime\prime}(0)$ of Eq.~\eqref{eq:taylor}. 
These  last two terms are essential to describe the $S$-factor in the Solar Gamow peak energy since they generate a $\sim 7$\% and $\sim 0.5$\% corrections respectively to $S(E)$ around $E=0$~\cite{SFIII}.
Higher derivative terms can be safely neglected, although their inclusion (at least the third derivative term) helps to stabilize the values for the lower derivatives~\cite{Acharya2016,Acharya2023}.

Historically, most of the theoretical efforts have been devoted to the prediction of $S(0)$. The many studies summarized in Refs.~\cite{SFI,SFII,SFIII} lead to the present recommended value of $S(0)=4.09(1\pm 0.015)\times 10^{-25}$  MeV b~\cite{SFIII}. The central value is obtained averaging three results: the first one is that of SFII~\cite{SFII}, obtained using the phenomenological highly-accurate Argonne $v_{18}$ potential~\cite{Wiringa1995} (AV18), and the nuclear axial current constructed within the phenomenological or the chiral effective field theory ($\chi$EFT) approach - the so-called hybrid $\chi$EFT in this last case. The other two values are those of Refs.~\cite{Acharya2023,DeLeon2023}, obtained, respectively, within a fully consistent $\chi$EFT approach, based on non-local chiral interactions and chiral axial currents, or within the so-called pionless effective field theory ($\slashed{\pi}$EFT) approach, where also the pion is considered as a heavy degree of freedom and therefore it is integrated out. In all these studies, the two-body axial current presents a coupling constant ($z_0(c_D)$ in $\chi$EFT, $L_{1,A}$ in $\slashed{\pi}$EFT) which is fixed to the Gamow-Teller matrix element in tritium $\beta$-decay, as proposed in the pioneer work of Ref.~\cite{Carlson1992}, and then systematically implemented in all calculations not only of the weak proton capture~\cite{Schiavilla1998,Marcucci2013,Marcucci2019,Acharya2016,Acharya2023,DeLeon2023}, but also for other reactions as the proton weak capture on $^3$He (the so-called $hep$ reaction)~\cite{Marcucci2000a,Marcucci2000b,Park2003}, and the muon capture on deuteron and light nuclei~\cite{Marcucci2011,Marcucci2012,Marcucci2018E,Gnech2024}. The error of 1.5\% assigned to $S(0)$ in Ref.~\cite{SFIII} is obtained  in the adopted averaging procedure between the $\chi$EFT, $\slashed{\pi}$EFT results and that of SFII, increased by 0.9\% to account for the update in the single-nucleon axial coupling constant $g_A$ between SFII and SFIII. In addition, a 1\% error was added in order to account for any input that would tend to move all results in a coordinated way, as, for example, it happened for the $g_A$ constant. 
The first and second derivatives $S^{\prime}(0)/S(0)$ and  $S^{\prime\prime}(0)/S(0)$ are obtained averaging the $\chi$EFT and $\slashed{\pi}$EFT predictions of Refs.~\cite{Acharya2023} and~\cite{Chen2013}, and the recommended values are $S^{\prime}(0)/S(0)=(11.0\pm 0.2)$ MeV$^{-1}$ and $S^{\prime\prime}(0)/S(0)=(242\pm 72)$ MeV$^{-2}$. To be noticed that, given the energy dependence of $S(E)$, the parametrization of Eq.\eqref{eq:taylor} and consequently the values of $S^{\prime}(0)/S(0)$ and  $S^{\prime\prime}(0)/S(0)$ are in principle not necessary.

The results summarized above and presented in SFIII maintain some open questions: (i) the result of SFII, obtained with the AV18 local potential, increased by 0.9\% as discussed above, is in agreement with the results obtained within $\chi$EFT and $\slashed\pi$EFT taking into account the theoretical errors. However, the AV18 value for $S(0)$ remains slightly lower than the other two (4.05 vs.\ 4.10 and 4.12 in $\chi$EFT and $\slashed\pi$EFT, respectively, in units of 10$^{-25}$ MeV b). Note that the calculations in $\chi$EFT have been performed using only non-local potential models~\cite{Marcucci2013,Acharya2016,Marcucci2019,Acharya2023}. (ii) In SFIII, the quantification of the uncertainties for $S(0)$ and its derivatives is based on direct comparison of different modeling, rather than statistical tools. The development of $\chi$EFT  permits in principle to overcome this problem, but a Bayesian analysis of the truncation errors of the chiral current and interaction expansions, and an estimate of error arising from model dependence has not been performed so far for the $pp$ reaction. 

In this work, for the first time,  we leverage on the most recent development in Bayesian analysis of truncation errors~\cite{Melendez2019} to compute and analyze the $pp$ capture rate and the theoretical errors in the context of $\chi$EFT using both local and non-local interactions.
We follow the procedure developed in Ref.~\cite{Gnech2024} for the muon capture, focusing  on theoretical error estimate
based on the Bayesian analysis of the truncation errors of the chiral current and interaction expansions, and on the model dependence.

The importance of a robust estimate for the theoretical uncertainty on $S(E)$ is evident, keeping in mind that the $pp$ fusion is the first reaction of the $pp$ chain, and therefore an uncertainty on $S(E)$ out of control could have significant consequences on the $pp$ Solar neutrino fluxes.  

\subsection{Overview of the present calculation}
\label{subsec:overview}
In this work we have adopted a large variety of highly-accurate nuclear chiral potentials, both local and non-local. In particular, we have focused on two different families of potentials, developed by two distinct research groups. The first family of potential are the so-called Entem-Machleidt-Nosyk (EMN) interactions, derived in Ref.~\cite{EMN2017}. These are $\Delta$-less non-local potentials, developed in momentum-space,  for which all the chiral orders from leading-order (LO) up to next-to-next-to-next-to leading order (N$^3$LO) are available. The interactions are regularized using a non local cutoff function characterized by a parameter $\Lambda$. Three different interactions with $\Lambda=$ 450, 500 and 550 MeV have been used at all orders up to N$^3$LO, all fitted on the nucleon-nucleon scattering data up to 300 MeV. Note that the EMN potentials are available also at N$^4$LO, but no consistent axial current exists, and therefore such models have not been included in this study.
 The second family of potentials are the so-called Norfolk (NV) potentials, derived in Ref.~\cite{Norfolk2016}. These are $\Delta$-full local potentials, developed in coordinate-space at fixed order (N$^3$LO) in the chiral expansion. The regularization of these interactions are made through two regulators, one for the short-range components ($R_S$) and the other one for the long-range terms ($R_L$). Four different interactions are available with two different cutoff sets, each of them fitted to the scattering data up to laboratory energies of 125 or 200 MeV. The list of all the interactions used with the relevant details are summarized in Table~\ref{tab:potlist}.
 \begin{table}
  \centering
  \renewcommand{\arraystretch}{1.5}
  \begin{tabular}{cccccc}
    \hline
    \hline
    Name & DOF & $O_\chi$ & $(R_{\rm S},R_{\rm L})$ or $\Lambda$ & $E$ range & Space \\
    \hline
    NVIa  & $\pi,N,\Delta$ & N$^3$LO & $(0.8,1.2)$ fm & 0--125 MeV & $r$\\
    NVIb  & $\pi,N,\Delta$ & N$^3$LO & $(0.7,1.0)$ fm & 0--125 MeV & $r$\\
    NVIIa & $\pi,N,\Delta$ & N$^3$LO & $(0.8,1.2)$ fm & 0--200 MeV & $r$\\
    NVIIb & $\pi,N,\Delta$ & N$^3$LO & $(0.7,1.0)$ fm & 0--200 MeV & $r$\\
    \hline
    EMN450     & $\pi,N$        & LO--N$^3$LO & $450$ MeV      & 0--300 MeV & $p$\\
    EMN500     & $\pi,N$        & LO--N$^3$LO & $500$ MeV      & 0--300 MeV & $p$\\
    EMN550     & $\pi,N$        & LO--N$^3$LO & $550$ MeV      & 0--300 MeV & $p$\\
    \hline
    \hline
  \end{tabular}
\caption{\label{tab:potlist}Summary of two-nucleon interactions used in this study. In the first column we indicate the name adopted to identify each interaction and in the remaining columns we list its main features, including degrees of freedom (DOF), chiral order ($O_\chi$), cutoff values ($(R_{\rm S},R_{\rm L})$ for local and $\Lambda$ for non-local potentials), laboratory energy range over which the fits to the two-nucleon database have been carried out ($E$ range), and whether it is  expressed in configuration ($r$) or in momentum ($p$) space.} \end{table}

To be remarked that the EMN potentials used in this work are of the same family of the non-local N$^3$LO500 and N$^3$LO600 potentials~\cite{Entem2003,Machleidt2011} used in previous calculations of the $pp$ reaction~\cite{Marcucci2013,Marcucci2019,Acharya2023}, and that NV potentials are very similar in the operatorial structure to the phenomenological AV18 potentials~\cite{Wiringa1995}. These similarities will permit us to perform a meaningful comparison with the previous calculation and they will allow us to find the reason of why the AV18 value for $S(0)$ remains slightly smaller than the values of Refs.~\cite{Marcucci2013,Marcucci2019,Acharya2016,Acharya2023}.

The only relevant transition for the $S$-factor around the Gamow peak is the one from $^1S_0$ $pp$ initial state and the $J^\pi=1^+$ deuteron state that is generated by the axial current operator. We use the axial currents  developed by the JLab-Pisa group in Ref.~\cite{Baroni2016} for the EMN interactions and in Ref.~\cite{Baroni2018} for the NV potentials up to N$^3$LO. Here we summarize the main contributions, providing the corresponding expression in Sec.~\ref{sec:nucl_curr}. At LO ($Q^{-3}$) the axial current has the one-body Gamow-Teller operator associated with the axial coupling constant $g_A=1.2723$~\cite{Patrignani2016}. At next-to-next-to leading order (N$^2$LO - $Q^{-1}$) the current receives contribution from the relativistic corrections (RC) of the Gamow-Teller operator and only for the NV interactions from the one-pion exchange with the intermediate excitation of a $\Delta$ (OPE-$\Delta$).
At N$^3$LO ($Q^0$) the axial current receives contribution from the one-pion exchange (OPE) and a contact term (CT) characterized by the low-energy constant (LEC) $z_0$ (see Eq.~\eqref{eq:jN3LO_CT} below).
Since the LEC $z_0$ is linearly dependent on the LEC $c_D$ (see Eq.~\eqref{eq:z0} and~\eqref{eq:z0_EMN} below) that appears in the three-nucleon interaction,
this has been determined fitting contemporary the $^3$H
binding energy and the Gamow-Teller matrix element of the $^3$H
$\beta$-decay. The value of $c_D$ (and $c_E$) used for the EMN and NV interactions can be found in Refs.~\cite{Gnech2024}  and~\cite{Baroni2018} respectively.

The calculation proceeds as follows. We calculate the deuteron wave function $\Psi_d^{M=\pm 1,0}$ and the $pp$ wave function in the $^1S_0$ channel $\Psi_{pp}^{^1S_0}$, using the chiral Hamiltonian for all the available models. The electromagnetic contributions beyond
the leading Coulomb interaction, such as two-photon exchange and vacuum polarization, are also retained in the $pp$ wave function, following the procedure of Ref.~\cite{Marcucci2013} for the non-local interactions (for the local interactions, such inclusion presents no difficulty).
Then the matrix element of the axial current consistent with the used interaction is calculated at  LO, N$^2$LO and N$^3$LO.
Using this matrix element, the cross section $\sigma(E)$ and the astrophysical $S$-factor $S(E)$ of Eq.~\eqref{eq:sfactor} are calculated using the formulas in Sec.~\ref{sec:details} in the range $E=3-30$ keV, at energy intervals of 1 keV. This choice for the energy range is based on two observations: (i) below 3 keV the numerical results become unstable, as already seen in Refs.~\cite{Marcucci2013,Acharya2023} and verified again in this work; (ii) the maximum energy of 30 keV
is enough to make stable the extraction of the zero-energy $S$-factor and its derivatives, when the Taylor expansion of Eq.~\eqref{eq:taylor} is used, as it has been shown in Refs.~\cite{Acharya2016,Acharya2023}.
The energy dependence of $S(E)$ has been parameterized as a third-degree Taylor polynomial, which parameters has been fitted to the calculated points with the minimum $\chi^2$ method. The values  of $S(0)$, $S^\prime(0)$ and $S^{\prime\prime}(0)$ obtained for the various potentials will be presented in Sec.~\ref{subsec:res}. The relative error per datum of the fitting procedure is $\sim 0.013$\%  and it is completely negligible compared to the other uncertainty sources. 

\subsection{Theoretical uncertainties estimate}
\label{subsec:theouncer}
The main sources of theoretical uncertainties are the truncation of the chiral expansion of the interactions and currents, and the one associated to the use of different nuclear interactions (model dependence). 
We calculate the uncertainty of the $S$-factor due to the expansion truncation using the Bayesian analysis method described in Refs.~\cite{PhysRevC.92.024005,Melendez2019}, and applied in Ref.~\cite{Gnech2024} to the muon capture process on deuteron. The model dependence uncertainty is instead estimated using the procedure outlined in Refs.~\cite{William2021,Gnech2024}.

Let us start with the truncation errors. Following the procedure of Ref.~\cite{Gnech2024} in our analysis we assume the chiral expansions of the current and interaction to be independent and we analyze them separately.
Therefore,  we fix the interaction order at N$^3$LO for both the EMN and the NV potentials (in fact the NV potentials are only available at this order) to study the truncation error of the current expansion.
Similarly, to study the interaction truncation error, we fix the current order at N$^3$LO, notice that the LEC $z_0$ is set to 0 when we use the interaction at LO and NLO. This last analysis is performed only using the EMN interactions for which all the chiral orders are available.

When we study the truncation of the nuclear interaction or the nuclear current, for a fixed interaction model of Table \ref{tab:potlist}, we assume that the $n$-th order energy dependent $S$-factor calculations can be factorized as~\cite{PhysRevC.92.024005,Melendez2019}
\begin{equation} \label{eq: S factor calcolated as sum NnLO}
    S^{(i)}_n(E) = S^{(i)}_{\text{ref}} (E)\sum_{k=0}^n c^{(i)}_k(E) \left(\frac{Q(E)}{\Lambda_\text{b}}\right)^{\nu_k^{(i)}} \: ,
\end{equation}
where $c^{(i)}_k(E)$ are dimensionless coefficients, $S^{(i)}_{\text{ref}} (E)$ is the reference dimension-full scale that is chosen such that the parameters $c^{(i)}_k(E)$ are of order 1,  $Q(E)$ is the expansion parameter defined as in Ref.~\cite{Melendez2019} to be
\begin{equation} \label{eq: Q value energy dependent error}
    Q(E) = \frac{(2m_pE)^4+m_\pi^8}{(2m_pE)^{7/2}+m_\pi^7} \: ,
\end{equation} 
and $\Lambda_\text{b}$ is the breakdown scale of the theory that for this work is assumed to be $\Lambda_\text{b}=550$ MeV. Further information regarding this choice can be found in Sec.~\ref{subsec:bayes2}. 
The index $i= \{ {\rm int, cur}\}$ indicates that the variable depends explicitly on the fact that we are considering the truncation of the interaction or the current, respectively. The exponent $\nu_k^{(i)}$ is the power counting. For the truncation error of the interaction $\nu_{0,1,2,3}^{(\text{int})}=\{0,2,3,4\}$, while for the truncation error of the currents $\nu_{0,1,2}^{(\text{cur})}=\{0,2,3\}$.
Notice that in this work we consider only the JLab-Pisa group power counting for the weak axial current.

The missing contribution to the theoretical prediction due to the chiral expansion truncation is then naturally defined as
\begin{equation}
  \delta S^{(i)}_n(E)=S^{(i)}_{\text{ref}} (E)\sum_{k=n+1}^{\infty}c^{(i)}_k(E)\left(\frac{Q(E)}{\Lambda_\text{b}}\right)^{\nu_k^{(i)}}\,.
  \label{eq:trunc_error}
\end{equation}
Our goal is then to determine this truncation error and the uncertainty associated
with it. The idea of Ref.~\cite{Melendez2019} is to build a stochastic representation of the $c^{(i)}_k(E)$
based on a Gaussian Process (GP) that emulates our order-by-order chiral calculation. The parameters of the GP are chosen using prior distributions that are updated during the training of the GP, using the Bayes theorem. The training of the GP is performed using a subset of the calculated points. The remaining points are used to perform statistical tests on the quality of the final GP, as shown in Sec.~\ref{subsec:bayes2}. 
Once validated, the GP is then exploited to emulate the missing $c^{(i)}_{k>n}(E)$ coefficients that appear in the $\chi$EFT truncation errors of Eq.~\eqref{eq:trunc_error}.
As a result, we obtain a distribution for the missing contribution given by
\begin{equation}
    \delta S^{(i)}_n(E) \sim \text{GP}\left[\mu_n^{(i)}(E), \sigma_n^{(i)}(E)\right]\ ,
    \label{eq:missing-distribution}
\end{equation}
where the mean value $\mu_n^{(i)}(E)$ is the truncation error and the standard deviation $\sigma_n^{(i)}(E)$ is the uncertainty associated with it. Therefore, the expectation value of the $S$-factor estimated truncating the chiral expansion at the $n$-th order is given by 
\begin{equation}
    \tilde{S}_n^{(i)}(E) = S_n^{(i)}(E) + \mu_n^{(i)}(E)\ ,
    \label{eq:s_tilde}
\end{equation} 
with an associated uncertainty $\sigma_n^{(i)}(E)$ .
To check the validity of the GP procedure, we will provide also an estimate of the truncation error and the uncertainty $\sigma_n^{(i)}(E)$ obtained using the prescription
of Ref.~\cite{Epelbaum2014}, described in Sec.~\ref{subsec:moredetails}. Note that this second prescription is similar to assuming a uniform prior for the error distribution.

The interaction truncation uncertainty in the $S$-factor calculated with the NV potentials cannot be determined using the method previously explained, since the NV potentials are fixed at N$^3$LO. In these cases, we considered the theoretical prediction as our best estimate, i.e. the calculation obtained with current and interaction at N$^3$LO, with the associated uncertainty estimated as the half-range uncertainty for all the NV potential models, i.e. 
\begin{equation}\label{error int NV}
    \sigma_{\text{N}^3\text{LO}}^{(\text{int})}(E) = \frac{|S_{\text{NVIa}}(E)-S_{\text{NVIIb}}(E)|}{2} \: ,
\end{equation}
since $S_{\text{NVIa}}(E)$ and $S_{\text{NVIIb}}(E)$ are, respectively, the maximum and minimum value obtained for the $S$-factor for the NV interactions in the energy range considered. 

The final expectation value and uncertainty associated to a specific nuclear interaction model is  calculated as the weighted arithmetic mean between the two best estimates at N$^3$LO for the interaction and the current truncation. Assuming the two distributions are independent and normally distributed with the same expectation value, the maximum likelihood method gives as estimator  of the mean 
\begin{equation} \label{eq: total theoretical prediction}
    \tilde{S}^{(\text{tot})} (E) = \sigma_{\text{trun}}^2 (E) \sum_i \frac{\tilde{S}_{\text{N}^3\text{LO}}^{(i)} (E)}{[{\sigma_{\text{N}^3\text{LO}}^{(i)}} (E)]^2} \: ,
\end{equation}
where the total truncation variance is defined as
\begin{equation} \label{eq: truncation error}
    \sigma_{\text{trun}}^2 (E) = \frac{1}{\sum_i{[\sigma_{\text{N}^3\text{LO}}^{(i)}} (E)]^{-2}} \:,
\end{equation}
which is again the best estimator for the variance.

Finally, to analyze the uncertainties due to the model dependence, we follow the approach proposed in Ref.~\cite{William2021} and used in Ref.~\cite{Gnech2024} for muon capture on deuteron. Using the best estimate $\tilde{S}^{(\text{tot})}(E)$, we obtain the average value of the energy-dependent astrophysical $S$-factor~\cite{William2021,Gnech2024} as
\begin{equation} \label{eq: Mean S final results}
    \langle S(E) \rangle = \sum_j \tilde{S}_{j}^{(\text{tot})}(E)\: P_j \: ,
\end{equation}
where $j$ denotes the various potential models used in this work (summarised in Table~\ref{tab:potlist}), and $P_j$ is the probability of choosing model $j$.  Since there is no \textit{a priori} reason to favor one model over another, and no reason to privilege local over non-local potentials, we assign $P_j=1/6$ for EMN potentials, and $P_j=1/8$ for NV ones, so that each family has a probability of $1/2$ to be chosen. We can write then the energy dependent uncertainty associated to $\langle S(E) \rangle$ as~\cite{William2021,Gnech2024}
\begin{equation} \label{variance S factor total}
    \sigma_S^2 (E) = \sigma_{\text{syst}}^2 (E) + \sum_j \sigma_{\text{trun},j}^2 (E) \: P_j \: ,
\end{equation}
where $\sigma_{\text{trun},j}(E)$ is the truncation uncertainty associated with the potential model $j$ in Eq.~\eqref{eq: truncation error}, and $\sigma_{\text{syst}}$ is the systematic uncertainty, expressed as
\begin{equation}
    \sigma_{\text{syst}}^2 (E) = \sum_j [\tilde{S}_{j}^{(\text{tot})}(E)]^2\: P_j - \left[\sum_j \tilde{S}_{j}^{(\text{tot})} (E)\: P_j\right]^2 \: .
\end{equation}

\subsection{Results}
\label{subsec:res}

In this section, we present the results for the astrophysical $S$-factor as a function of the center-of-mass energy $E$, as well as the values for $S(0)$ and its first and second derivative $S^\prime(0)$ and $S^{\prime\prime}(0)$. 

\begin{figure}[thb]
    \includegraphics[width=0.48\textwidth]{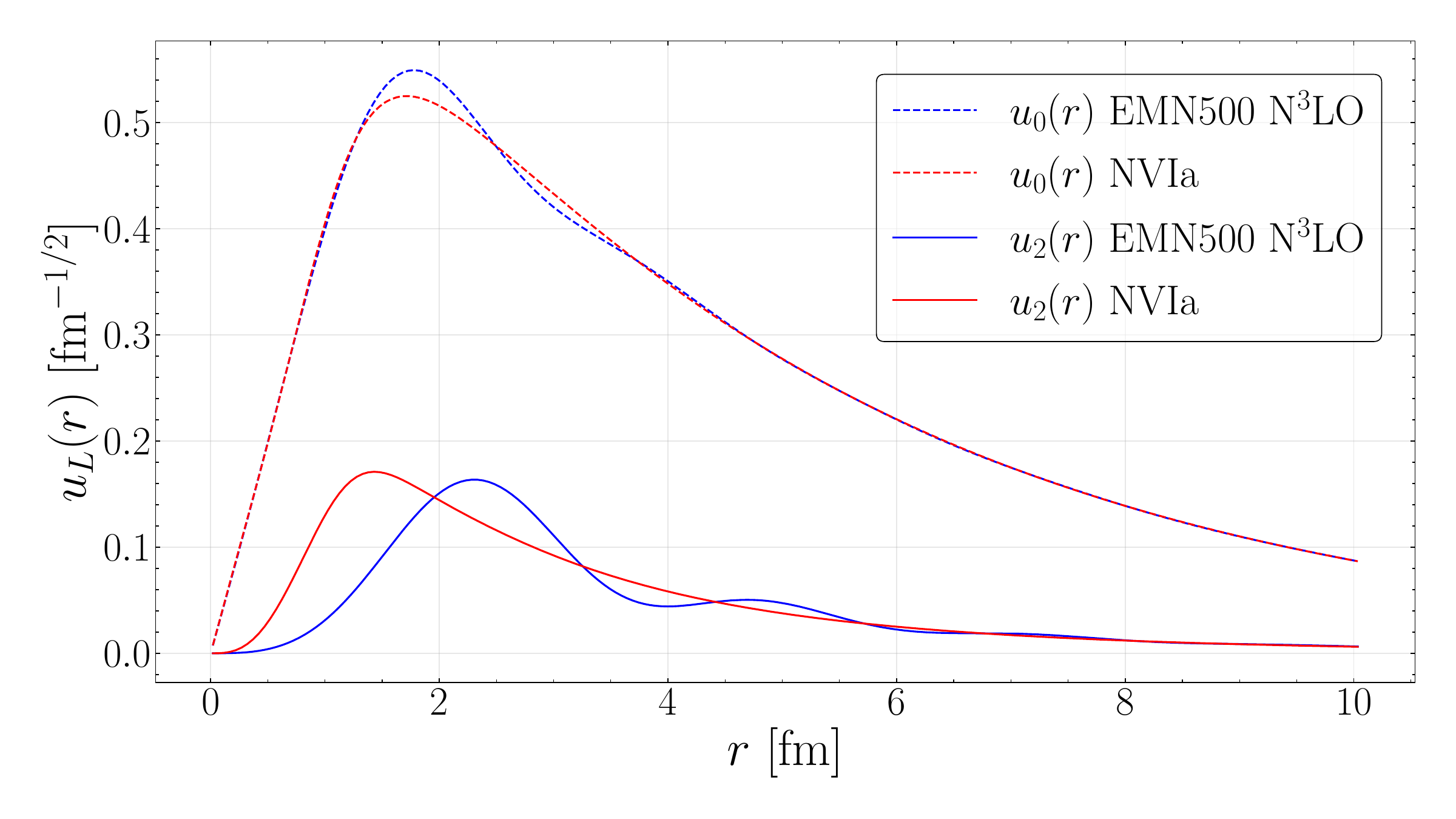}
    \caption{The $s$-wave $u_0(r)$ (dashed lines) and $d$-wave $u_2(r)$ (solid lines) reduced radial wave functions calculated using the
NVIa potential (red lines) and the EMN potential at N$^3$LO and cutoff value $\Lambda$ = 500 MeV (blue lines).\label{fig:deuteron}}
\end{figure}

In Table~\ref{tab:s0_int} we present the values of $S(0)$ varying the order of the EMN potentials and using the currents at N$^3$LO with $z_0=0$ for the LO and NLO interactions and $z_0$ fixed to reproduce the tritium $\beta$-decay for the rest. As it can be seen the results shows a good convergence behavior. In Table~\ref{tab:s0_cur} we present instead the values of $S(0)$ for the NV and the EMN potentials at N$^3$LO for all the chiral orders of the current. Also in this case the convergence of the chiral expansion is reasonable. As it can be seen from Table~\ref{tab:s0_cur}, the values for $S(0)$ obtained with for each order of the current using the local NV potentials are systematically smaller than the values obtained with the EMN non-local interactions. This has been traced back to the shapes of the deuteron wave function components. As it can be seen from  Fig.~\ref{fig:deuteron}, the deuteron wave functions obtained with the non-local EMN500 and the local NVIa potentials are indeed quite different. Similar behaviors are present for the other local and non-local interactions. The wave functions, especially the $d$-wave component, computed with the EMN500 potential are shifted to larger values of the relative distance $r$ with respect to those calculated with the NV potentials in coordinate space. Moreover, the EMN wave functions present a wiggling behavior that is not present when working with local potentials.
Very likely this behavior can be ascribed to the adopted regularization function in EMN potentials, and it was already noticed with older version of the non-local chiral potentials (see for instance Fig.~16 of Ref.~\cite{Machleidt2011}). This effect is very likely the origin of the differences between the results for $S(0)$ obtained with the local phenomenological AV18 potential~\cite{Schiavilla1998,Park2003} and used in SFII~\cite{SFII}, and those obtained with the non-local chiral potentials~\cite{Marcucci2013,Marcucci2013,Marcucci2019,Acharya2016,Acharya2023}, and used in SFIII~\cite{SFIII}.

\begin{table}[hbt]
\renewcommand{\arraystretch}{1.5}
\begin{tabular}{ccccc}
\hline\hline
Model & $S_{\rm LO}^{(\text{int})}(0)$ & $S_{\rm NLO}^{(\text{int})}(0)$ &  $S_{\rm N^2LO}^{(\text{int})}(0)$ &  $S_{\rm N^3LO}^{(\text{int})}(0)$
\\
\hline
EMN450 & 4.087 & 4.089 & 4.099 & 4.090 \\
EMN500 & 4.119 & 4.083 & 4.098 & 4.091 \\
EMN550 & 4.152 & 4.063 & 4.100 & 4.089 \\
\hline\hline
\end{tabular}
\caption{$S$-factor parameter $S(0)$ in $10^{-25}$ MeV b, calculated with the EMN potentials from LO up to N$^3$LO using the current at N$^3$LO.}\label{tab:s0_int}
\end{table}

\begin{table}[htb]
\renewcommand{\arraystretch}{1.5}
\begin{tabular}{cccc}
\hline\hline
Model & $S_{\rm LO}^{(\text{cur})}(0)$ &  $S_{\rm N^2LO}^{(\text{cur})}(0)$ &  $S_{\rm N^3LO}^{(\text{cur})}(0)$
\\
\hline
NVIa & 4.018 & 4.071 & 4.062 \\
NVIb & 4.014 & 4.093 & 4.061 \\
NVIIa & 3.990 & 4.043 & 4.036 \\
NVIIb & 3.986 & 4.064 & 4.030 \\
\hline
EMN450 & 4.035 & 4.031 & 4.090 \\
EMN500 & 4.050 & 4.045 & 4.091 \\
EMN550 & 4.056 & 4.051 & 4.089 \\
\hline\hline
\end{tabular}
\caption{$S$-factor parameter $S(0)$ in $10^{-25}$ MeV b, calculated using the NV and EMN potentials at N$^3$LO. The current order runs from LO up to N$^3$LO.}\label{tab:s0_cur}
\end{table}

\begin{figure}[thb]
    \includegraphics[width=0.48\textwidth]{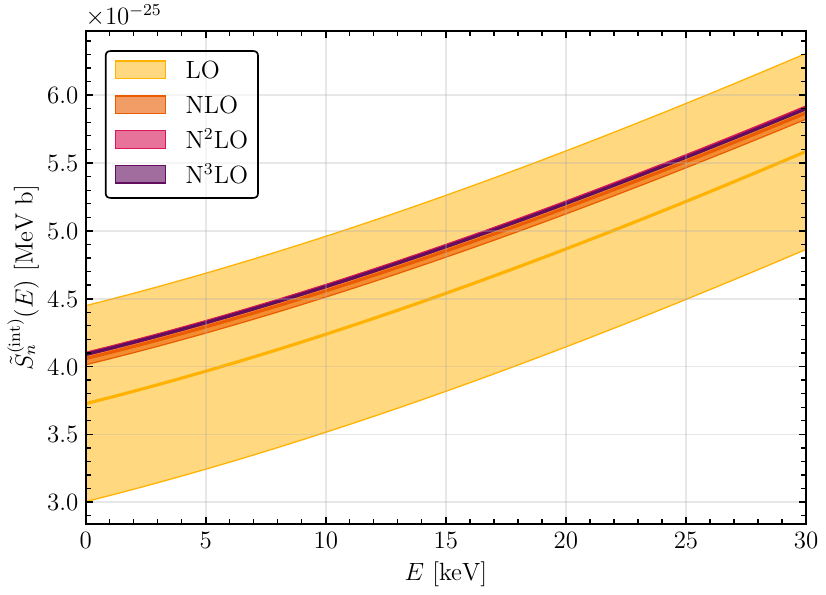}
    \caption{The astrophysical $S$-factor expectation value ${\tilde S}_n^{({\rm int})}(E)$ calculated using the EMN potential with fixed cutoff $\Lambda$ = 500 MeV and with the current at N$^3$LO with $z_0$ fixed at zero for the LO and NLO potentials, and fixed to reproduce the tritium $\beta$-decay in the other cases. Each band color represents the $\sigma^{(\text{int})}_n(E)$ truncation uncertainty at each interaction order: yellow band for LO ($n = 0$), orange band for NLO ($n = 1$), red band for N$^2$LO ($n = 2$) and purple band for N$^3$LO ($n = 3$).} \label{fig:se_pot}
\end{figure}

In Fig.~\ref{fig:se_pot} we present the astrophysical $S$-factor expectation value ${\tilde S}_n^{({\rm int})}(E)$ (see Eq.~\eqref{eq:s_tilde}) for the EMN500 interaction together with the truncation error $\sigma^{({\text{int}})}_n(E)$ from $n=0$ (LO) up to $n=3$ (N$^3$LO) in the energy interval $0-30$ keV. The current is fixed at N$^3$LO with $z_0$ fixed at zero for the LO and NLO potentials, and fixed to reproduce the tritium $\beta$-decay in the other cases. Each bandwidth represents the uncertainty for a given order due to the truncation of the interaction chiral expansion. Note that, as the order increases, the error decreases. Furthermore, each band is fully compatible with the band of the previous order, showing a good convergence of the chiral expansion of the astrophysical $S$-factor as the order of the interaction increases. Note that, because the curve corresponds to very low energies, the momentum scale $Q(E)$ defined in Eq.~\eqref{eq: Q value energy dependent error} remains constant, essentially $Q(E) \simeq m_{\pi}$. As a result, the widths of the uncertainty bands are constant over the energy range considered.
Similar results are obtained using the other EMN potentials.

\begin{figure}[thb]
    \includegraphics[width=0.48\textwidth]{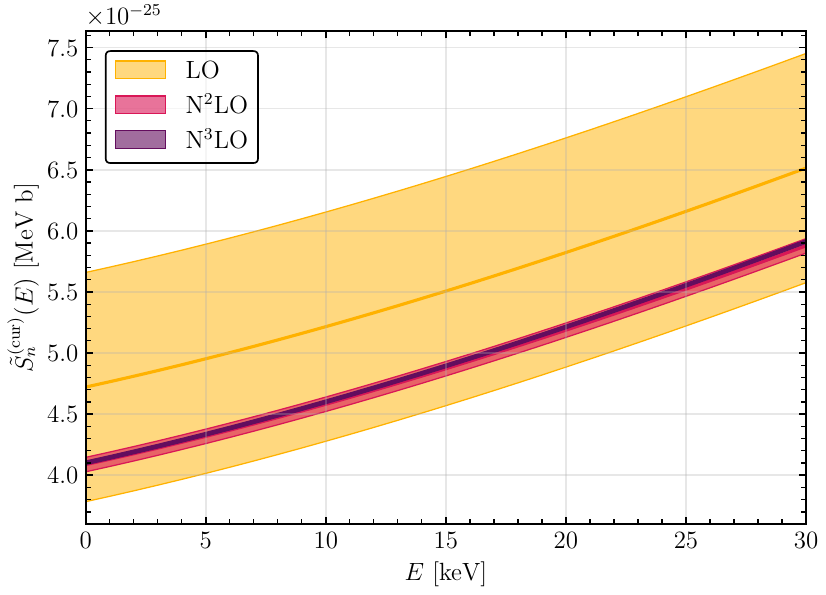}
    \caption{The astrophysical $S$-factor expectation value ${\tilde S}_n^{({\rm cur})}(E)$ calculated using the EMN potential at N$^3$LO with fixed cutoff $\Lambda$ = 500 MeV. Each band color represents the $\sigma^{(\text{cur})}_n(E)$ truncation uncertainty at each current order: yellow band for LO ($n = 0$), red band for N$^2$LO ($n = 1$) and purple band for N$^3$LO ($n = 2$).} \label{fig:se_cur}
\end{figure}

In Fig.~\ref{fig:se_cur} we present the astrophysical $S$-factor expectation value ${\tilde{S}}_n^{({\text{cur}})}(E)$ (see Eq.~\eqref{eq:s_tilde}) for the EMN500 interaction, varying the cutoff order, together with the truncation error $\sigma^{({\text{cur}})}_n(E)$. In this case, the index $n$ can assume the values $0, 1, 2$, corresponding to LO, N$^2$LO, and N$^3$LO, respectively. 
Each bandwidth represents the uncertainty for a given current order. Again, as the order increases,  the error decreases and the results are compatible with the estimate at the previous order showing a nice convergence.  Similar results are obtained for the other EMN potentials as well as for the NV potentials. 

We summarized the $S(0)$ parameter prediction prediction $\tilde{S}_{\text{N}^3\text{LO}}^{(i)}(0)$ for the interaction and current chiral expansion truncation and the relative uncertainties $\sigma_{\text{N}^3\text{LO}}^{(i)}(0)$ in Table~\ref{tab:s0_bayes} for each potential model considered. The total prediction and the relative truncation variance in the fourth column are calculated using Eqs.~\eqref{eq: total theoretical prediction} and \eqref{eq: truncation error}. Notice how the central values $\tilde{S}_{\text{N}^3\text{LO}}^{(\text{int})}(0)$ and $\tilde{S}_{\text{N}^3\text{LO}}^{(\text{cur})}(0)$ are different in particular for the EMN interactions but still compatible within $1\sigma$. Notice also that the errors associated with the truncation of the currents is in general larger. This is due to the correction generated at N$^3$LO by the $z_0$ that, because of the fitting procedure, reabsorbs effects also beyond its actual order and then does not follow exactly the power expansion.

\begin{table}[htb]
\renewcommand{\arraystretch}{1.5}
\begin{tabular}{cccc}
\hline\hline
Model & $\tilde{S}_{\rm N^3LO}^{(\text{int})}(0)$ &  $\tilde{S}_{\rm N^3LO}^{(\text{cur})}(0)$ &  $\tilde{S}^{(\text{tot})}(0)$
\\
\hline
NVIa & 4.062(16) & 4.063(14) & 4.063(11) \\
NVIb & 4.061(16) & 4.059(28) & 4.060(14) \\
NVIIa & 4.036(16) & 4.037(13) & 4.037(10) \\
NVIIb & 4.030(16) & 4.027(30) & 4.029(14) \\
\hline
EMN450 & 4.088(3) & 4.103(17) & 4.088(3) \\
EMN500 & 4.089(3) & 4.101(15) & 4.090(3) \\
EMN550 & 4.087(4) & 4.098(13) & 4.088(4) \\
\hline\hline
\end{tabular}
\caption{Theoretical prediction of the $S$-factor parameter $S(0)$ in $10^{-25}$ MeV b. In the second and third columns the estimation truncating the interaction and current chiral expansion at N$^3$LO respectively, for all the potential models. In the fourth column the total theoretical prediction and the overall truncation uncertainty. In parentheses we report the Bayesian error.}\label{tab:s0_bayes}
\end{table}

\begin{figure}[thb]
    \includegraphics[width=0.48\textwidth]{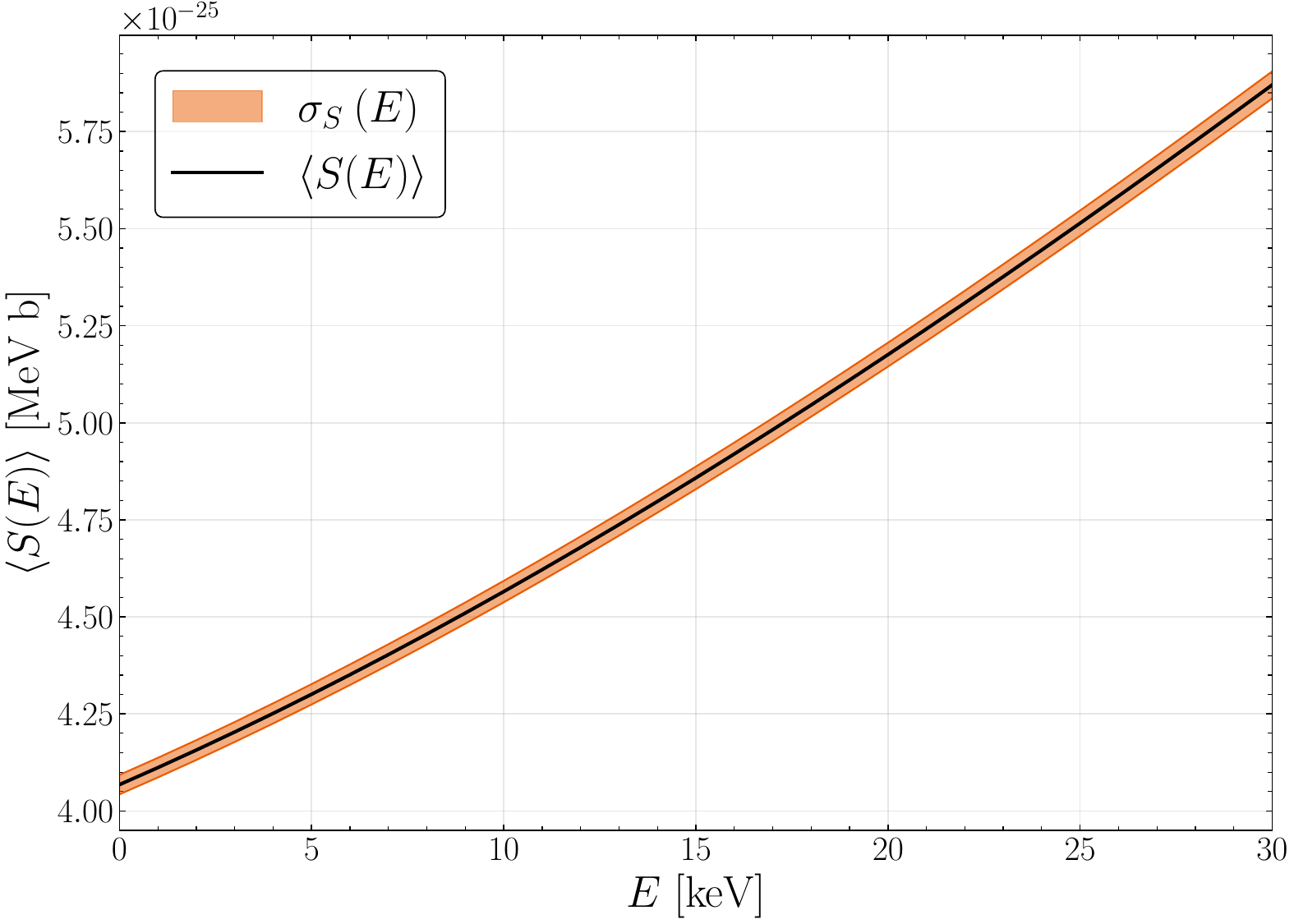}
    \caption{The average value of the energy-dependent $S$-factor $\langle S(E)\rangle$, as defined in Eq.~\eqref{eq: Mean S final results} (black line) together with the energy-dependent variance $\sigma_S (E)$ as defined in Eq.~\eqref{variance S factor total} (orange band).}
\label{fig:se_tot}
\end{figure}

Following the procedure outlined at the end of Sec.~\ref{subsec:theouncer}, we present in Fig.~\ref{fig:se_tot} the average value for $S(E)$ in the range $0-30$ keV, together with the energy dependent 1$\sigma$ uncertainty given in Eq.~\eqref{variance S factor total}. The resulting uncertainty band is narrow  being the error of the order of $0.6\%$ over the entire energy interval.
The average value and the theoretical uncertainty for  $S(0)$ calculated in this way is
\begin{equation}
     \begin{split}
         \langle S(0)\rangle &= 4.068 \pm 0.025 \times 10^{-25} \:\text{MeV} \,\,\, \text{b} \\
        &= 4.068(1 \pm 0.006) \times 10^{-25} \:\text{MeV} \,\,\, \text{b} \: \ .
    \label{eq:S-final-Bayes}
    \end{split}
\end{equation}

We compare our result with the recent SFIII review~\cite{SFIII}, which quotes $S(0)=4.09\:(1\pm0.015)\times10^{-25}\:\text{MeV b}$. The uncertainty quoted in Ref.~\cite{SFIII} includes a 1\% contribution accounting for $g_A$ error, in addition to a 0.5\% uncertainty associated with the truncation of the chiral expansion. Our uncertainty is derived within a Bayesian framework considering both the truncation error of currents and interactions as well as a statistical estimate of the model dependence and thus represents a statistically meaningful and robust estimate. Our result is compatible within 1$\sigma$ with the SFIII recommendation even if our central value is slightly lower. This small shift can be attributed to the wider set of nuclear interactions considered here, which includes both local and non-local chiral potentials, whereas previous studies in the framework of $\chi$EFT were restricted to non-local interactions.  We also find excellent agreement between our results obtained with the EMN potentials and previous calculations based exclusively on non-local chiral interactions~\cite{Marcucci2013,Marcucci2019,Acharya2023}. Similarly a good agreement is obtained when comparing the AV18 result, $S(0)=4.05\times10^{-25}\:\text{MeV b}$~\cite{Schiavilla1998}, with the corresponding values obtained using the NV interactions. This shows once again the effect of the local nature of the interaction on $S(0)$.

In order to obtain an additional estimate of the error, we computed the truncation errors on the expansion of the current  and interactions also by following the procedure of Ref.~\cite{Epelbaum2014}. In this case, the parameter $S(0)$ and its theoretical uncertainty are given by
\begin{equation}
\begin{split}
     \langle S^{\text{\scriptsize{\cite{Epelbaum2014}}}}(0) \rangle &= 4.069 \pm 0.031 \times 10^{-25} \:\text{MeV} \,\,\, \text{b} \\
    &= 4.069(1 \pm 0.008) \times 10^{-25} \:\text{MeV} \,\,\, \text{b} \: \ .
\label{eq:S-final-Epelbaum}
\end{split}
\end{equation}
The error estimate is consistent with that obtained with the Bayesian approach and shown in Eq.~\eqref{eq:S-final-Bayes}.
Further information on this second procedure can be found in Sec.~\ref{subsec:moredetails}.

Finally, we fitted the $S$-factor using the Taylor expansion in Eq.~\eqref{eq:taylor}, considering our estimated uncertainty $\sigma_S(E)$. In Table~\ref{tab:resultlist} we report the results of $S'(0)$ and $S''(0)$, where we compare the present results with those of SFIII~\cite{SFIII}. The quoted uncertainties  once propagated permit to reconstruct the error band shown in Figure~\ref{fig:se_tot} that contains both the truncation errors and the model dependence. 
As can be seen from the Table~\ref{tab:resultlist}, the uncertainties on the derivatives are very small.
On the other hand, the errors of Ref.~\cite{SFIII} are much larger, likely due to the fact that they are obtained considering both $\chi$EFT and $\slashed{\pi}$EFT approaches.
\begin{table}[thb]
  \centering
  \renewcommand{\arraystretch}{1.5}
  \setlength{\tabcolsep}{7pt} 
  \begin{tabular}{c|cc}
    \hline
    \hline
    $S(E)$ Parameter & This work & SFIII \cite{SFIII} \\
    \hline
    $ S(0) $ [$10^{-25}$ MeV b] & $4.068\pm0.025$ & $4.090\pm0.061$ \\
    $ S'(0)/ S(0) \: [\text{MeV}^{-1}]$ & $10.60\pm0.01$ & $11.0\pm0.2$ \\
    $ S''(0)/ S(0) \: [\text{MeV}^{-2}]$ & $347.9\pm0.9$ & $242\pm72$ \\
    \hline
    \hline
  \end{tabular}
\caption{\label{tab:resultlist} Estimate $S$-factor parameters $S(0)$, $S'(0)$ and $S''(0)$ and the relative error results comparison between this work and SFIII.}
\end{table}

\subsection{Conclusions and astrophysical implications}
\label{subsec:concl}

In summary, we have presented a new study of the $S(E)$ astrophysical $S$‑factor within $\chi$EFT, employing a wide range of nuclear potential models and consistent weak currents, and for the first time considering both local and non‑local interactions. In particular, we have considered the local NV potentials, which share a similar operatorial structure with the phenomenological AV18 interaction used in previous studies~\cite{Schiavilla1998,Marcucci2000a,Marcucci2000b,Park2003} and forming the basis for the SFII recommended value~\cite{SFII}, as well as the non‑local EMN interactions from LO up to N$^3$LO for several cutoff values $\Lambda$. These are the most recent versions of the N$^3$LO500 and N$^3$LO600 potentials employed in earlier calculations~\cite{Marcucci2013,Marcucci2019,Acharya2023} and forming the basis of the SFIII recommendation~\cite{SFIII}.

For the first time in this context, we have performed a Bayesian estimate of the theoretical uncertainty associated with the truncation of the chiral expansion of interactions and currents. The resulting value of $S(0)$ reported in Eq.~\eqref{eq:S-final-Bayes} carries an uncertainty of about 0.6\%. Its central value is lower than the SFIII recommendation by roughly 0.5\%, and it is in good agreement with SFII. We attribute this difference to the systematic discrepancy between results obtained with local NV and non‑local EMN potentials, which can be further traced to differences in the short‑range structure of the corresponding deuteron wave functions (see Fig.~\ref{fig:deuteron}).

We then investigated the astrophysical implications of this new determination, focusing on stellar evolutionary models and Solar neutrino fluxes. To this end, {we considered three values of $S(0)$ (in units of $10^{-25}$ MeV b), i.e.\ 4.068, 3.993, and 4.143, corresponding respectively to the Bayesian central value of Eq.~\eqref{eq:S-final-Bayes}, and to the values lower and higher than the central value by 3$\sigma$, with $\sigma=0.025$ being the absolute error given also in Eq.~\eqref{eq:S-final-Bayes}.
Since stellar models depend on several physical inputs affected by sizable uncertainties, the key issue is whether such a variation in $S(0)$ produces observable effects beyond existing model uncertainties.

We first considered age determinations of stellar clusters, which rely on the luminosity at the central hydrogen exhaustion point, identified with the turn‑off (TO) or overall contraction (OC) phase. Following standard practice, we focused on the bright turn‑off (BTO) luminosity for old clusters, defined as the point that is $\sim$0.1 dex more luminous and about 100 K cooler than the TO~\cite{Chaboyer96}. Using linear sensitivity estimates from the literature~\cite{Valle13}, we found that even considering a 3$\sigma$ error, thus a $\sim 2$\% variation of $S(0)$, the change in the BTO luminosity is about $2-2.5$\textperthousand, an effect lower than the ones due to the other main uncertainty sources~\cite{Valle13}.
A similarly effect is found for younger, metal‑rich open clusters, where the OC luminosity is most sensitive to $S(0)$ for ages between 2 and 4 Gyr~\cite{Tognelli15}. In this case, taking into account a 3$\sigma$ uncertainty in $S(0)$, the maximum luminosity variation is slightly below 1\%, a very small effect that has a minor relevance in cluster age determination compared to other uncertainty sources~\cite{Chaboyer96, Chaboyer98, Krauss03, Valle13, DeglInnocenti09}.

We finally examined the impact of the new $S(0)$ estimate within the framework of the Standard Solar Model. Since the Solar luminosity is fixed by observations, variations in the $pp$ $S$‑factor are compensated by changes in the central temperature $T_c$. Using established scaling relations~\cite{Villante21}, we find that a $1\sigma$ error in $S(0)$ leads to a change in $T_c$ of less than 1\textperthousand. When a $3\sigma$ uncertainty in $S(0)$ is considered, the corresponding change in $T_c$ is slightly less than 2\textperthousand. These results are consistent with full Solar model calculations~\cite{Tognelli15}. As a consequence, helioseismic quantities and surface abundances remain essentially unaffected.
The largest effects are found in the predicted neutrino fluxes from temperature‑sensitive reactions. In particular, the fluxes of $^8$B, $^{15}$O, and $^{17}$F neutrinos vary by at most slightly less than $\sim$2\% at 1$\sigma$ and $\sim$5\% at 3$\sigma$, while the $pp$, $pep$, and $hep$ fluxes remain nearly unchanged due to their strong luminosity constraint~\cite{Bahcall89, Serenelli13, Villante21}. A variation of $\sim$5\% of the neutrino fluxes is not, in principle, negligible. However, it should be noted that this result is obtained by assuming a $3\sigma$ uncertainty in $S(0)$, a very conservative estimate. Other Solar model inputs, when varied within the same error range, have a larger impact on neutrino fluxes~\cite{Villante21}.
In conclusion, a 3$\sigma$ variation in $S(0)$ has a negligible impact on stellar age determinations and induces at most few‑percent changes in Solar neutrino fluxes. A substantially more precise determination of $S(E)$ would become relevant only if uncertainties in other theoretical inputs were comparably reduced, and if future neutrino measurements, particularly for the $^8$B, $^{15}$O, and $^{17}$F components, would achieve percent‑level accuracy.

\section{Details of the calculation}
\label{sec:details}

\subsection{Formalism}
\label{subsec:formalism}
The $S$-factor of Eq.~\eqref{eq:sfactor} is obtained from the $pp$ weak capture energy dependent cross section $\sigma(E)$, which in the center-of-mass (CM) reference frame can be written as
\begin{equation} \label{eq:defcrosssection}
\begin{aligned}
    \sigma (E) 
    &= \sqrt{\frac{\mu}{2E}} \int \frac{d\bm{p}_e}{(2\pi)^3}\frac{d\bm{p}_{\nu}}{(2\pi)^3} F(Z,E_e) \: \overline{|T_{\text{fi}}|^2} \: \\
    &\qquad \times \: 2\pi \delta \left(\Delta m + E - \frac{q^2}{2m_d} - E_e - E_\nu\right) \: \: ,
\end{aligned}
\end{equation}
where $E$ is the CM energy, $\mu = m_p/2$ is the reduced mass, $\bm{p}_e$ ($\bm{p}_\nu$) and $E_e$ ($E_\nu$) are the electron (neutrino) momentum and energy, $\Delta m=2m_p-m_d$, with $m_p$ and $m_d$ the proton and deuteron masses, and $\bm{q}=\bm{p}_e+\bm{p}_\nu$ is the momentum transfer. The Coulomb interaction between the outgoing positron and the final deuteron, including radiative corrections \cite{Kurylov_2003}, is taken into account thanks the Fermi function $F(Z,E_e)$, with $Z=1$ being the charge of the deuteron.
In Eq.~\eqref{eq:defcrosssection} $\overline{|T_{\text{fi}}|^2}$ is the probability density function between the initial and the final states, summed over the final state spins and mediated over the initial ones, i.e.\
\begin{equation} \label{eq: probability density summed over spin}
    \overline{|T_{\text{fi}}|^2} \equiv \frac{1}{4} \sum_{s_1,s_2} \sum_{s_d,s_e,s_\nu} |\braket{f|{H}_W|i}|^2 \: .
\end{equation}
We have indicated with $s_1$ and $s_2$ the spins of the two protons, and $s_d$, $s_e$, $s_\nu$ the deuteron, positron and neutrino spins. In Eq.~\eqref{eq: probability density summed over spin} ${H}_W$ is the weak interaction Hamiltonian operator, which under the assumption that the weak interaction is a contact interaction can be written as~\cite{Marcucci2000b}
\begin{equation} \label{eq: weak hamiltonian}
    {H}_W = \frac{G_V}{\sqrt{2}} \int d\bm{x}  \: \ell^{\mu} (\bm{x}) h_{\mu} (\bm{x}) \: ,
\end{equation}
where $\ell^{\mu} (\bm{x})$ and $h_{\mu} (\bm{x})$ are the the leptonic and hadronic four-vector current densities, and $G_V$ is the so-called vector Fermi coupling constant, $G_V=1.14939 \times 10^{-5} \: \text{GeV}^{-2}$~\cite{HARDY1990429}. Using the definition of Eq.~\eqref{eq: weak hamiltonian} the weak interaction Hamiltonian matrix element can be rewritten as
\begin{equation} \label{eq: matrix element compacted}
    \braket{f|{H}_W|i} = \frac{G_V}{\sqrt{2}}  \: \ell^{\mu}\: \braket{\Psi_{d} (\bm{p}_d,s_d)|h_{\mu}^\dagger (\bm{q})|\Psi_{pp} (\bm{p}, s_1, s_2)} \: ,
\end{equation}
where $\ell^\mu =  \overline{u}_\nu \gamma^\mu (1-\gamma_5) v_e$ is the leptonic current, $\overline{u}_\nu=u_\nu^\dagger \gamma^0$, $u_\nu$ and $v_e$ being the antineutrino and electron Dirac spinors, $\Psi_{d} (\bm{p}_d, s_d)=e^{i{\bm p}_d\cdot {\bm R_{\text{CM}}}} \Psi_d({\bm r},s_d)$ and $\Psi_{pp} (\bm{p}, s_1, s_2)$ are the deuteron and the proton-proton wave functions, with $\bm{p}$ the proton-proton relative momentum, and $h^{\mu} (\bm{q})$ is the Fourier transform of the hadronic current density whose spatial part is presented in Sec.~\ref{sec:nucl_curr}.

The deuteron wave function in coordinate space is written as
\begin{equation} \label{deuteron wave function}
    \Psi_d(\bm{r},s_d) = \sum_{L=0,2} R_L(r) \left[Y_L(\hat{\bm{r}}) \otimes \chi_{S=1}\right]_{1s_d} \xi_{00} \: ,
\end{equation}
where $\bm{r}$ is the relative distance between the neutron and the proton, $\left[Y_L(\hat{\bm{r}}) \otimes \chi_{S}\right]_{JJ_z}$ is the angular part of the wave function, $\xi_{TT_z}$ is the isospin component with $T,T_z=0,0$, and $R_L(r)$ is the radial part of the wave function, obtained numerically using the Rayleigh-Ritz variational principle (see Ref.~\cite{Marcucci2011}). 

The proton-proton wave function can be decomposed in a partial wave expansion as
\begin{eqnarray}
\label{eq: proton-proton wave function partial wave exp}
    \Psi_{pp} (\bm{p}, s_1, s_2) &=& 4\pi \sum_{S,S_z}  \braket{\frac{1}{2} \: s_1, \frac{1}{2} \: s_2 | S \: S_z} \nonumber \\
    &\times&\sum_{LL_zJJ_z} i^L Y_{LL_z}^* (\hat{\bm{p}}) \nonumber \\ &\times& \braket{S \: S_z, L \: L_z|J \: J_z} \: \overline{\Psi}_{pp}^{LSJJ_z} (p) \: ,
\end{eqnarray}
where $J$ and $J_z$ ($L$ and $L_z$, $S$ and $S_z$) represent the total angular momentum (orbital angular momentum, total spin) and its third component, respectively. Furthermore in Eq.~\eqref{eq: proton-proton wave function partial wave exp} $Y_{LL_z}^* (\hat{\bm{p}})$ is the complex conjugate of the spherical harmonic, and $\overline{\Psi}_{pp}^{LSJJ_z} (p)$ is the proton-proton wave function with unitary flux, computed numerically by using the Kohn variational principle (see Ref.~\cite{Marcucci2013}). As shown in Refs.~\cite{Acharya2016,Marcucci2019,Acharya2023}, in the low-energy regime the contribution to $S(E)$ of partial waves above ${}^1S_0$ is less than 1\%. Therefore they can be considered negligible, in order to make calculations simpler and less time consuming. 

\subsection{Explicit expression of the nuclear axial current}\label{sec:nucl_curr}

At the energy of interest for this work only the axial weak current operators are relevant. 
Here we report the various operators in coordinate space obtained in the chiral expansion by the JLab-Pisa group~\cite{Baroni2016,Baroni2018}. We do not include here the pion-pole contributions, which are negligible at the considered low energy regime. The operators reported here are consistent with the ones derived by the Bochum group even if using a different power counting (see Ref.~\cite{Krebs2020} for a recent review). 

The LO axial current is defined by the one-body axial current operator and it is written as~\cite{Baroni2018}
\begin{equation} \label{eq:jLO}
{\bm j}_{\text{LO}}({\bm q}) = -\frac{g_A}{2} \tau_{i,\pm}{\bm \sigma}_i {\rm e}^{{\rm i}{\bm q}\cdot{\bm r}_i} + (i \leftrightarrow j)\ ,    
\end{equation}
where $g_A=1.2723$ is the single-nucleon axial coupling constant~\cite{Patrignani2016}, ${\bm \sigma}_i$ is the spin Pauli matrix of the $i$-th nucleon, $\tau_{i,\pm} = (\tau_{i,x} \pm {\rm i} \tau_{i,y})/2$, $\tau_{i,x/y}$ is the Pauli $x/y$ isospin matrix, ${\bm q}$ the momentum
transfer, and ${\bm r}_i$ the position of the $i$-th nucleon.
The N$^2$LO contributions to the axial current are generated by relativistic correction (RC)
to the LO term and the one-pion exchange (OPE) term with $\Delta$-isobar intermediate excitation. The OPE contribution is not included with the EMN potentials, being $\Delta$-less models. The N${}^2$LO axial current is therefore written as
\begin{equation}\label{eq:jN2LO}
    {\bm j}_{\text{N$^2$LO}}({\bm q}) = {\bm j}_{\text{RC}}({\bm q}) + {\bm j}_\Delta({\bm q}) \ ,
\end{equation}
where
\begin{eqnarray}
  {\bm j}_{\text{RC}}({\bm q})=&-&\frac{g_A}{4m^2} \tau_{i,\pm}
  {\rm e}^{i{\bm q}\cdot{\bm r}_i}
  \Big[{\bm \sigma}_i \nabla^2 -\nabla_i ({\bm \sigma}_i\cdot{\bm \nabla}_i)\nonumber \\
    &-&\frac{q^2}{4} {\bm \sigma}_i - \frac{1}{2}({\bm q}\times{\bm \nabla}_i) + i \Big({\bm \sigma}_i ({\bm q}\cdot{\bm \nabla}_i)
    \nonumber \\
    &-&
    \frac{1}{2}{\bm q}({\bm \sigma}_i \cdot {\bm \nabla}_i)
    -\frac{1}{2} ({\bm \sigma}_i\cdot{\bm q}){\bm \nabla}_i\Big) \Big]
  \nonumber \\
  &+& (i\leftrightarrow j)
  \label{eq:jN2LO_RC} \ ,
  \end{eqnarray}
and
\begin{eqnarray}
    {\bm j}_\Delta({\bm q}) = &-&{\rm e}^{i{\bm q}\cdot {\bm r}_i} ({\bm \tau}_i \times {\bm \tau}_j )_{\pm} [I^{(1)}(\mu_{ij};\alpha_1^\Delta) {\bm \sigma}_i \times {\bm \sigma}_j \nonumber \\
    &+& I^{(2)}(\mu_{ij}; \alpha_1^\Delta) {\bm \sigma}_i \times \hat {\bm r}_{ij} {\bm \sigma}_j \cdot \hat {\bm r}_{ij}] \nonumber \\
   &-& {\rm e}^{i{\bm q}\cdot {\bm r}_i} \tau_{j,\pm} [I^{(1)} (\mu_{ij}; \alpha_2^\Delta) {\bm \sigma}_j \nonumber \\
   &+& I^{(2)} (\mu_{ij}; \alpha_2^\Delta ) \hat {\bm r}_{ij} {\bm \sigma}_j \cdot \hat {\bm r}_{ij}] + (i \leftrightarrow j)\ .
\label{eq:jN2LO_Delta}
\end{eqnarray}
Note that in this work we corrected a bug in the code associated with the RC contribution. The error was associated with the term proportional to $q^2$ and therefore it has a minimal impact on the previously computed results related to low-energy nuclear reactions.
 Finally, at N$^3$LO contributions to the axial current are given by the $\Delta$-less OPE and by a contact term (CT), so that
\begin{equation}
    {\bm j}_{\text{N$^3$LO}}({\bm q})={\bm j}_{\text{OPE}}({\bm q}) + {\bm j}_{\text{CT}}({\bm q}) \ ,
    \label{eq:jN3LO}
\end{equation}
where
\begin{eqnarray}
    {\bm j}_{\text{OPE}}({\bm q})= &-&{\rm e}^{i{\bm q}\cdot {\bm r}_i} ({\bm \tau}_i \times {\bm \tau}_j )_{\pm} [I^{(1)}(\mu_{ij};\alpha_1) {\bm \sigma}_i \times {\bm \sigma}_j \nonumber \\
    &+& I^{(2)}(\mu_{ij}; \alpha_1) {\bm \sigma}_i \times \hat {\bm r}_{ij} {\bm \sigma}_j \cdot \hat {\bm r}_{ij}] \nonumber \\
   &-& {\rm e}^{i{\bm q}\cdot{\bm r}_i} \tau_{j,\pm} [I^{(1)} (\mu_{ij}; \alpha_2) {\bm \sigma}_j\nonumber \\
   &+& I^{(2)} (\mu_{ij}; \alpha_2 )\hat {\bm r}_{ij} {\bm \sigma}_j \cdot \hat {\bm r}_{ij}] 
   \nonumber \\
   &-&\frac{1}{2m} ({\bm \tau}_i\times{\bm \tau}_j)_{\pm}\nonumber \\
   &\times&\left\{ {\bm p}_i\, , \, {\rm e}^{i{\bm q}\cdot{\bm r}_i} {\tilde I^{(1)}}(\mu_{ij};{\tilde{\alpha}}_1) {\bm \sigma}_j\cdot {\hat{\bm r}}_{ij} \right\}\nonumber \\
   &-& \frac{i}{m} ({\bm \tau}_i\times{\bm \tau}_j)_{\pm} {\rm e}^{i{\bm q}\cdot{\bm r}_i} {\tilde{I}}^{(1)}(\mu_{ij};{\tilde \alpha}_2)\nonumber \\
   &\times & ({\bm \sigma}_i\times {\bm q}) {\bm \sigma}_j\cdot{\hat{\bm r}}_{ij} + (i \leftrightarrow j)\ ,
\label{eq:jN3LO_OPE}
\end{eqnarray}
where ${\bm r}_{ij} = {\bm r}_i - {\bm r}_j$,  $\mu_{ij}= m_\pi r_{ij}$ for the NV interactions with $m_\pi$ being the pion mass,  $\mu_{ij}=\Lambda r_{ij}$ for the EMN interactions with $\Lambda$ the interaction cutoff,  $\left\{\cdots,\cdots\right\}$ the anticommutator,  ${\bm p}_i=-i\nabla_{r_i}$, and $m$ the nucleon mass.
The contact term reads
\begin{eqnarray}
{\bm j}_{\text{CT}}({\bm q})=z_0
 {\rm e}^{i{\bm q}\cdot{\bm R}_{ij}} I^{(\text{CT})}(z_{ij})
  ({\bm \tau}_{i}\times {\bm \tau}_j)_a ({\bm{\sigma}}_i\times {\bm \sigma}_j) \label{eq:jN3LO_CT}\ ,
\end{eqnarray}
where ${\bm R}_{ij}= ({\bm r}_i+{\bm r}_j)/2$, and $z_{ij} = r_{ij} /R_S$ for the NV interactions with $R_S$ the short range cutoff, while $z_{ij} = \Lambda r_{ij} $ for the EMN interactions with $\Lambda$ the corresponding cutoff. The values of $R_S$ or $\Lambda$ used for each interaction are reported in Table~\ref{tab:potlist}.

In Eqs.~\eqref{eq:jN2LO_Delta} and~\eqref{eq:jN3LO_OPE}, the $I$-functions are known as correlation functions. For the NV local interactions they are given by
\begin{align}
    I^{(1)}(\mu,\alpha)&=-\alpha(1+\mu)\frac{{\rm e}^{-\mu}}{\mu^3}\,,\\
    I^{(2)}(\mu,\alpha)&=\alpha(3+3\mu+\mu^2)\frac{{\rm e}^{-\mu}}{\mu^3}\,,\\
    \tilde I^{(1)}(\mu,\tilde\alpha)&=-\tilde\alpha(1+\mu)\frac{{\rm e}^{-\mu}}{\mu^2}\,,
\end{align}
where
\begin{align}
    \alpha_1^\Delta&=\frac{g_A}{8\pi}\frac{m_\pi^3}{f_\pi^2}c_4^\Delta\,,\quad \alpha_2^\Delta=\frac{g_A}{4\pi}\frac{m_\pi^3}{f_\pi^2}c_3^\Delta\,,\\
      \alpha_1&=\frac{g_A}{8\pi}\frac{m_\pi^3}{f_\pi^2}\left(c_4+\frac{1}{4m}\right)\,,\quad \alpha_2=\frac{g_A}{4\pi}\frac{m_\pi^3}{f_\pi^2}c_3\,,\\
          \tilde\alpha_1&=\frac{g_A}{16\pi}\frac{m_\pi^2}{f_\pi^2}\,,\quad \tilde\alpha_2=\frac{g_A}{32\pi}\frac{m_\pi^2}{f_\pi^2}(c_6+1)\,,
\end{align}
with $f_\pi=92.4$ MeV being the pion decay constant. 
The regularization of the correlation functions is done consistently with the nuclear interactions, multiplying them for a local configuration-space cutoff function
\begin{equation}
    C_{R_L}(r)=1-\frac{1}{(r/R_L)^6 e^{2(r-R_L)/R_L}+1}\,,
\end{equation}
where the values of $R_L$ can be found in Table~\ref{tab:potlist} for each NV interaction used in this work.
The correlation functions corresponding to the EMN interactions are regularized in the momentum space  through a cutoff
\begin{equation}
    C_\Lambda(k)=\exp\left(-\frac{k^4}{\Lambda^4}\right)\,,
\end{equation}
where the values of $\Lambda$ can be found in Table~\ref{tab:potlist}. For these, the correlation functions appearing in Eq.~\eqref{eq:jN3LO_OPE} read
\begin{align}
    I^{(1)}(\mu,\alpha)&=-\frac{\alpha}{\mu}\int_0^\infty dx\frac{x^3}{x^2+\overline{m}_\pi^2}j_1(x\mu)C_\Lambda(x)\,,\label{eq:I1_EMN}\\
    I^{(2)}(\mu,\alpha)&=\alpha\int_0^\infty dx\,\frac{x^4}{x^2+\overline{m}_\pi^2}j_2(x\mu)C_\Lambda(x)\,,\\
    \tilde I^{(1)}(\mu,\alpha)&=-\tilde\alpha\int_0^\infty dx\frac{x^3}{x^2+\overline{m}_\pi^2}j_1(x\mu)C_\Lambda(x)\,,
\end{align}
where $\overline{m}_\pi=m_\pi/\Lambda$, and
\begin{align}
      \alpha_1&=\frac{g_A}{4\pi^2}\frac{\Lambda^3}{f_\pi^2}\left(c_4+\frac{1}{4m}\right)\,,\quad \alpha_2=\frac{g_A}{2\pi^2}\frac{\Lambda^3}{f_\pi^2}c_3\,,\\
  \tilde\alpha_1&=\frac{g_A}{8\pi^2}\frac{\Lambda^2}{f_\pi^2}\,,\quad \tilde\alpha_2=\frac{g_A}{16\pi^2}\frac{\Lambda^2}{f_\pi^2}(c_6+1)\,.
\end{align}

For the NV interactions the axial contact term in Eq.~\eqref{eq:jN3LO_CT} is regularized through a Gaussian cutoff 
\begin{equation}
I^{(\text{CT})}(z)=\frac{{\rm e}^{-z^2}}{\pi^{3/2}} \ ,
\end{equation}
and the adimensional LEC $z_0$ is given by~\cite{Baroni2018}
\begin{equation}
\begin{aligned}
z_0=\:&\frac{g_A m_\pi^2}{2 f_\pi^2}\frac{1}{(m_\pi R_S)^3} \Bigg[-\frac{m_\pi}{4g_A\Lambda_\chi}c_D\\ &+\frac{m_\pi}{3}(c_3+c_3^\Delta+2c_4+2c_4^\Delta) +\frac{m_\pi}{6 m} \Bigg]\ ,
\label{eq:z0}
\end{aligned}
\end{equation}
where $\Lambda_\chi=1$ GeV is the chiral symmetry breakdown scale.
For the EMN interaction the regularization function reads
\begin{equation}
    I^{(\text{CT})}(z)=\frac{1}{2\pi^2}\int_0^\infty dx \,x^2 C_\Lambda(x)j_0(xz) \ ,
\end{equation}
while the adimensional LEC $z_0$ is given by
\begin{equation}
\begin{aligned}
z_0=&\frac{g_A \Lambda^2}{ f_\pi^2} \Bigg[-\frac{\Lambda}{4g_A\Lambda_\chi}c_D \Bigg]\ .
\label{eq:z0_EMN}
\end{aligned}
\end{equation}
Notice that in this case the LECs $c_3$ and $c_4$, and the factor $1/m$ does not appear in the definition of $z_0$. This definition of $z_0$ is convenient for maintaining the same form of Eq.~\eqref{eq:jN3LO_OPE} for both the EMN and NV interactions. By rearranging Eq.~\eqref{eq:I1_EMN} it is possible also in the case of the EMN interactions to reabsorb part of the OPE terms in $z_0$ operator and obtain the same definition of $z_0$ shown in Eq.~\eqref{eq:z0}.

The values of the LECs $c_3$, $c_4$ and $c_6$ are
  fixed consistently with the adopted nuclear potential models. The LEC $c_D$, together with the well known LEC $c_E$ entering the three-nucleon interaction, is fixed with the procedure of Refs.~\cite{Marcucci2012,Gnech2024}. The adopted values are given in Ref.~\cite{Baroni2018} for the NV potentials, and in Table VII of Ref.~\cite{Gnech2024} for the EMN potentials.

\subsection{Additional details on the Bayesian analysis}\label{subsec:bayes2}
In this section we discuss the details of the Bayesian analysis performed to study the truncation errors of the 
chiral expansion of the interactions and the currents. We performed the analysis using a modified version of the \texttt{gsum} package~\cite{Melendez2019} already used in the Bayesian analysis of the muon capture in Ref.~\cite{Gnech2024}.

The analysis of the interaction expansion has been performed only for the EMN interactions. The expansion used for the interaction is defined in Eq.~\eqref{eq: S factor calcolated as sum NnLO} where $S_{\text{ref},j}^{(\text{int})}(E)= S^{(\text{int})}_{\text{LO},j}(0)$ (LO of the interaction and current fixed at N$^3$LO). The data set of $S(E)$ values consists of 31 points, that go from 0 to 30 keV. We separate this data set in a training set consisting of 3 points at energies $E=3,15,29$ keV and a test set consisting of 6 points at energies $E=5,9,13,17,21,25$ keV. More data points in the validation data set give rise to ill defined co-variance matrices. Finally, we use 0.00013  as the variance for the white noise in order to stabilize the matrix inversion (i.e.\ the nugget). In Fig.~\ref{fig:trunc_int}(a) we report the values for the coefficients $c_n^{(\text{int})}(E)$ obtained using the EMN450 model. The relatively linear behavior of the coefficients as a function of $E$ is related to the small energy range used in this analysis. As it can be seen, the GP prediction (shaded area) is indistinguishable from the $\chi$EFT calculation (full lines). 

In order to have a better understanding of the GP, we performed some statistical tests on the compatibility of the GP with our $\chi$EFT prediction provided by the \texttt{gsum} package: the Mahalanobis distance, the Pivoted Cholesky and the credible interval diagnostic. In Fig.~\ref{fig:trunc_int}(b) we report the Mahalanobis distance that measures the distance between the GP and the actual $\chi$EFT results. As it can be seen from the figure, the N$^2$LO result is compatible within the $95\%$ credible interval of the reference $\chi^2$ distribution (the whiskers in the figure), and the N${}^3$LO is compatible within the $68\%$ credible interval (the box in the figure). Furthermore, the results for the Mahalanobis distance are towards the minimal edge, indicating a very small distance between the emulated GP and the test points.
This coincidence is even more evident in the pivoted Cholesky diagnostics, shown in Fig.~\ref{fig:trunc_int}(c). Almost all test points lie within $1\sigma$, indicating that the GP predictions are statistically very close to the test points.
Finally, the credible interval diagnostic shown in Fig.~\ref{fig:trunc_int}(d) tests whether the truncation error computed at each order is compatible with the correction at the next order within a certain confidence level (CL). 
For example, the orange line in Fig.~\ref{fig:trunc_int}(d) is the credible interval for $\Delta S_2^{(\text{int})}(E)=S^{(\text{int})}_2(E)-S^{(\text{int})}_3(E)$ compared with the reference distribution of $\delta S^{(\text{int})}_2(E)$. As it can be seen from the figure, the orange line lays in the gray area showing a reasonable statistical behavior of the estimated error. However, the strong vertical behavior of the line indicates that the emulation is in general under-confident. In other words, the error predicted by the GP is  very large compared to the difference between the N$^2$LO and N$^3$LO. That is understandable since the differences among the simulated $S$-factor at N$^2$LO and N$^3$LO is minimal, as it can be seen in Fig.~\ref{fig:se_pot}. A similar conclusion can be drawn for $\Delta S_1^{(\text{int})}(E)=S^{(\text{int})}_1(E)-S^{(\text{int})}_2(E)$, corresponding to the purple line in Fig.~\ref{fig:trunc_int}(d).
The results obtained for all the other interaction models are qualitative very similar.

The analysis of the current expansion has been performed for both EMN and NV potential families. The reference dimension-full scale of Eq.~\eqref{eq: S factor calcolated as sum NnLO} is chosen to be $S_{\text{ref},j}^{(\text{cur})}(E)= S^{(\text{cur})}_{\text{LO},j}(0) (Q(E)/\Lambda_{\text{b}})^{-3}$ (LO of the current and interaction fixed at N$^3$LO). The $(Q(E)/\Lambda_{\text{b}})^{-3}$ is added to have $\nu_{k}^{{\rm (cur)}}$ starting from $0$. The same training and test points, and nugget used in the interaction case have also been used for this analysis. In Fig.~\ref{fig:trunc_cur} we report the results for the current truncation analysis using the EMN450 model. Also in this case the linear behavior of the coefficients shown in Fig.~\ref{fig:trunc_cur}(a) is due to the small energy range considered here. The Mahalanobis distance in Fig.~\ref{fig:trunc_cur}(b) indicates a high coincidence between the GP and the testing points, in particular for the N$^2$LO. Indeed only the N$^3$LO result is compatible within the $95\%$ credible interval. The same coincidence is also found in the pivoted Cholesky diagnostic, shown in
Fig.~\ref{fig:trunc_cur}(c), in which almost all test points are within 1$\sigma$. Finally, in Fig.~\ref{fig:trunc_cur}(d) the credible interval for the $\Delta S_2^{(\text{cur})}$ is shown, with a behavior similar to the one in the interaction analysis. All other interaction models yield qualitatively similar results.

The breakdown scale $\Lambda_\text{b}$ entering Eq.~\eqref{eq: S factor calcolated as sum NnLO} has been studied for each of the seven potential models considered in this work (see Table \ref{tab:potlist}). The optimal breakdown scale $\Lambda_{\text{b},j}^{\text{best}}$ for each potential model $j$ has been estimated as the one that maximizes the posterior distribution
$\text{pr}(\Lambda_\text{b} \mid \boldsymbol{E} \:, S_j(\boldsymbol{E}))$
introduced in Ref.~\cite{Melendez2019}, where $\boldsymbol{E}$ denotes the set of energy points used in the analysis and $S_j(\boldsymbol{E})$ the fitted $S$-factor for the $j$-th potential model evaluated at those energies. The results are reported in Table \ref{tab: lambda_best}.
\begin{table}[hbt]
\renewcommand{\arraystretch}{1.5}
\begin{tabular}{ccccc}
\hline\hline
Model & $\Lambda_{\text{b},j}^{\text{best}}$ [MeV]\\
\hline
NVIa & 750 \\
NVIb & 500 \\
NVIIa & 800 \\
NVIIb & 475 \\
\hline
EMN450 & 450 \\
EMN500 & 475 \\
EMN550 & 475 \\
\hline\hline
\end{tabular}
\caption{The optimal breakdown
scale for each potential model considered in this work, as obtained from the maximum of the posterior distribution.} \label{tab: lambda_best}
\end{table}
Note that for NVIa and NVIIa the breakdown scale is roughly equivalent to the mass of the $\rho$ meson, whereas for the other potential models the values are less than 500 MeV. The overall breakdown scale $\Lambda_\text{b}=550$ MeV used in this work has been obtained as an average value, similarly to the average $\langle S(E)\rangle$ given in Eq.~\eqref{eq: Mean S final results}.

\begin{figure*}[t]
  \centering
  \begin{subfigure}{0.25\textwidth}
    \includegraphics[width=\linewidth]{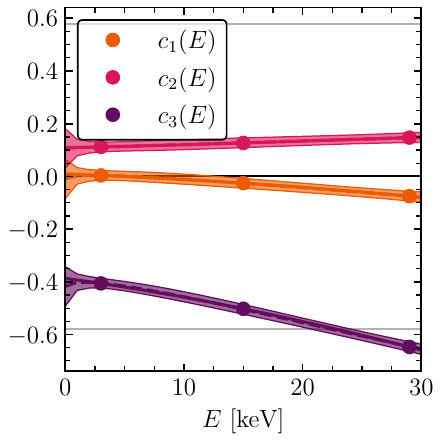}
    \caption{}
  \end{subfigure}
  \begin{subfigure}{0.16\textwidth}
    \includegraphics[width=\linewidth]{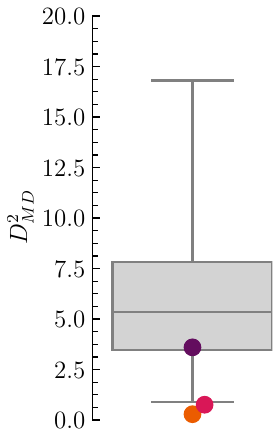}
    \caption{}
  \end{subfigure}
  \begin{subfigure}{0.25\textwidth}
    \includegraphics[width=\linewidth]{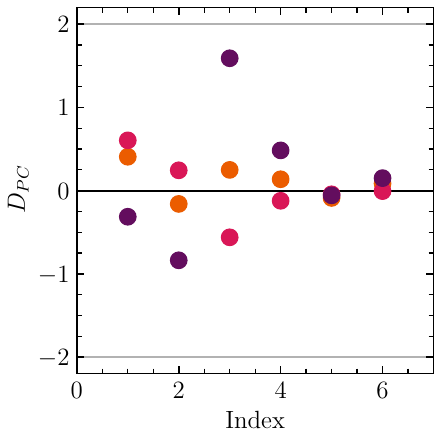}
    \caption{}
  \end{subfigure}
  \begin{subfigure}{0.25\textwidth}
    \includegraphics[width=\linewidth]{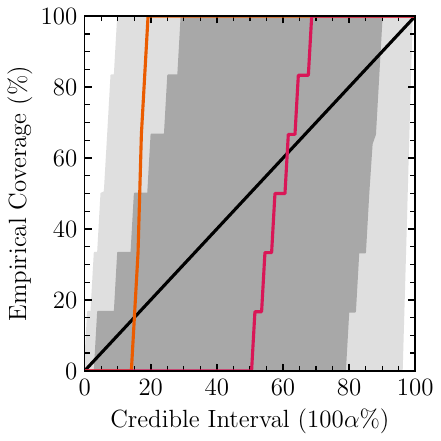}
    \caption{}
  \end{subfigure}
  \caption{\label{fig:trunc_int}The GP modeling of the interaction chiral expansion coefficients and its diagnostics for the EMN450 interaction. Panel (a): the simulators (solid lines - i.e. our calculation) along with the corresponding GP emulators (dashed lines) and their $2\sigma$ intervals (bands). The data used for training are denoted by dots. Panel (b): the Mahalanobis distances compared to the mean (interior line), $50\%$ (box) and $95\%$ (whiskers) credible intervals of the reference distribution. Panel (c): the pivoted Cholesky diagnostics versus the index along with $95\%$ credible intervals (gray lines). Panel (d): the credible interval diagnostics for the truncation error bands. The $1(2)\sigma$ is represented with the dark(light) gray band.}
\end{figure*}

\begin{figure*}[t]
  \centering
  \begin{subfigure}{0.25\textwidth}
    \includegraphics[width=\linewidth]{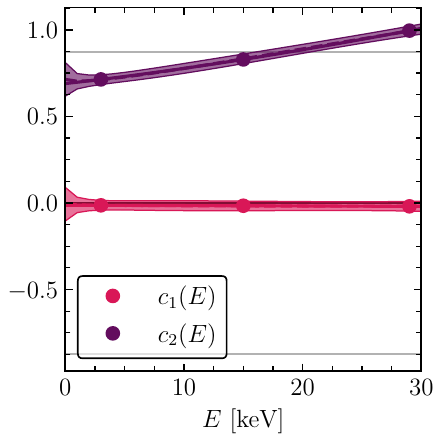}
    \caption{}
  \end{subfigure}
  \begin{subfigure}{0.16\textwidth}
    \includegraphics[width=\linewidth]{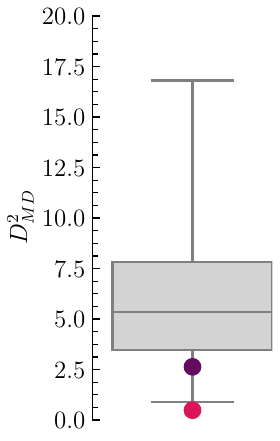}
    \caption{}
  \end{subfigure}
  \begin{subfigure}{0.25\textwidth}
    \includegraphics[width=\linewidth]{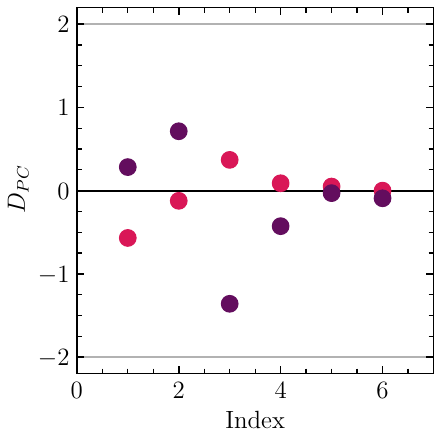}
    \caption{}
  \end{subfigure}
  \begin{subfigure}{0.25\textwidth}
    \includegraphics[width=\linewidth]{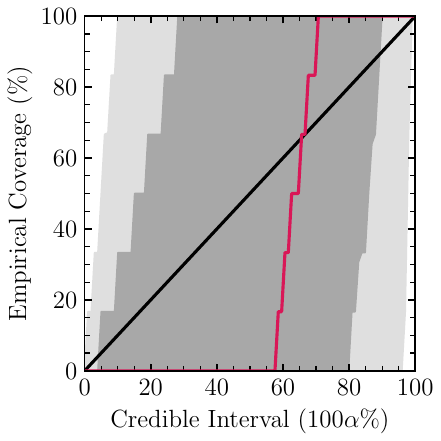}
    \caption{}
  \end{subfigure}
  \caption{\label{fig:trunc_cur}The same as Fig.~\ref{fig:trunc_int} but for the current chiral expansion using the EMN450 interaction.}
\end{figure*}

\subsection{Additional detailed results}
\label{subsec:moredetails}

An alternative approach to estimate the error arising from the order-by-order $\chi$EFT expansion has been proposed in Ref.~\cite{Epelbaum2014}. Following this procedure, the error arising from stopping at $n$-th order in the chiral expansion is estimated as
\begin{equation} \label{eq:error-a-la-Epelbaum}
\begin{split}
    \sigma_n^{(i)}(E) = \max_{k=0,\ldots, n} \Bigg\{ 
    &|S_k^{(i)}(E)-S_{k-1}^{(i)}(E)| \\
    &\times \left(\frac{Q(E)}{\Lambda_\text{b}}\right)^{\nu_n^{(i)}+1-\nu_k^{(i)}}
    \Bigg\} \, . \\
    &
\end{split}
\end{equation}
where the index $i=\{ \text{int},\text{cur} \}$ indicates which expansion we are varying, while fixing the other one at N${}^3$LO, the exponent $\nu_k^{(i)}$ is the power counting and $S_{-1}^{(i)}(E)=0$. The expansion parameter $Q(E)$ considered is the same of Eq.~\eqref{eq: Q value energy dependent error} and the breakdown scale value is the one derived in Sec.~\ref{subsec:bayes2}, i.e.\ $\Lambda_\text{b} = 550$ MeV.

This procedure has been used to estimate the theoretical error in Eq.~\eqref{eq:S-final-Epelbaum}. Note that in this case the mean value has been calculated as
\begin{equation}
    \langle S^{\text{\scriptsize{\cite{Epelbaum2014}}}}(E) \rangle = \sum_j S_{j} (E)\: P_j \:,
     \label{eq: mean value epelbaum}
\end{equation}
where $S_{j} (E)$ is the energy dependent $S$-factor fitted on the calculation obtained by fixing both the interaction and the current at N${}^3$LO.

\section*{Acknowledgments}
The work of V.B. is supported by the Deutsche Forschungsgemeinschaft (DFG) through the Collaborative Research Center 1660 "Hadron and Nuclei as discovery tools" (Project ID 514321794).

A.G.\ is supported by the Nuclear Theory for New Physics Topical Collaboration supported by the U.S.~Department of Energy under contract DE-SC0023663. A.G.\ acknowledge also the support of Jefferson Lab supported by the U.S. Department of Energy under contract DE-AC05-06OR23177. 

\bibliography{biblio.bib}
\bibliographystyle{apsrev4-1}
\end{document}